\documentclass[11pt]{article}
\usepackage{graphicx}
\begin{document}

\begin{center}
{\bf\Large Josephson effect in the cuprates:\\
microscopic implications}\\
\vspace{0.3cm}
R. Hlubina\\
{\it Department of Solid State Physics, Comenius University,\\
Mlynsk\'{a} dolina F2, SK-842 48 Bratislava, Slovakia}
\end{center}

\begin{abstract}
In the tunnel limit, the current-phase relation of Josephson junctions
can be expanded as $I(\phi)=I_1\sin\phi+I_2\sin 2\phi$.  Standard BCS
theory predicts that $I_1R_N\sim\Delta/e$ and $I_2/I_1\sim D$, where
$R_N$ is the resistance of the junction in the normal state, $\Delta$
is the superconducting gap, and $D\ll 1$ is the junction
transparency. In the cuprates, the experimental value of $I_1R_N$
($I_2/I_1$) is much smaller (larger) than the BCS prediction. We argue
that both peculiarities of the cuprates can be explained by
postulating quantum fluctuations of the pairing symmetry.
\end{abstract}

\section{Introduction}

The proximity to the metal-insulator transition is well known to lead
to an anomalous normal state of the high $T_c$ superconductors. It is
therefore interesting to ask whether, apart from the $d$-wave symmetry
of pairing, also the superconducting state of the cuprates can be
regarded as unconventional.  Most studies attempt to answer this
question by considering the properties of {\it quasiparticle
excitations}.  For instance, photoemission experiments seem to support
the conventional alternative, since sharp spectral functions have been
observed at low temperatures \cite{Damascelli02}.  On the other hand,
the effect of strong correlations on the {\it condensate} has remained
largely unexplored so far. This is surprising, since large quantum
phase fluctuations are to be expected in the cuprates as a combined
effect of the suppression of charge fluctuations and of the
uncertainty principle. In particular, such fluctuations have been
suggested to stabilize the RVB state and the presumably related
pseudogap phase of the cuprates \cite{Anderson97}. In this paper we
argue that by a detailed analysis of the Josephson effect, new
insights into the nature of quantum fluctuations in the cuprates can
be obtained.

Josephson junctions involving the cuprate superconductors have been
studied mainly because they enable phase-sensitive tests of pairing
symmetry \cite{Sigrist92}. Attention has been paid especially to two
particular types of Josephson junctions: grain boundary
\cite{Hilgenkamp02} and intrinsic \cite{Kleiner94} Josephson
junctions.  The best studied type of grain boundary Josephson
junctions involves junctions built on $c$-axis oriented films, where
the weak link forms at the boundary of grains which are rotated around
the [001] axis with respect to each other. Idealized junctions of this
type are characterized by a planar interface and two angles $\theta_1$
and $\theta_2$ between the interface normal and the crystallographic
directions in the grains forming the junction. The properties of the
junctions depend dominantly on the misorientation angle
$\theta=\theta_2-\theta_1$.  It is well known \cite{Hilgenkamp02} that
the transparency $D$ of grain boundary junctions decays exponentially
with increasing misorientation angle $\theta$, $D(\theta)\propto
\exp(-\theta/\theta_0)$. Thus, for $\theta>\theta_0\approx 5^\circ$,
grain boundary junctions are in the tunnel limit and (for $\theta$ not
too close to $45^\circ$, see Section~3) their current-phase relation
$I(\phi)=I_1\sin\phi+I_2\sin 2\phi+\ldots$ can be well approximated by
$I(\phi)=I_1\sin\phi$, neglecting the higher-order harmonics.

\section{The Josephson product}

A useful quantity characterizing the superconducting electrodes
forming the Josephson junction is the product of the first harmonic
$I_1$ with the junction resistance in the normal state,
$R_N$. According to standard theory (for homogeneous featureless
barriers), this so-called Josephson product is independent of the
junction area and of the barrier transparency, thus giving an
intrinsic information about the superconducting banks.  It is well
known that at interfaces between $d$-wave superconductors, anomalous
bound Andreev levels may form \cite{Hu94}.  At temperatures larger
than the energy of such Andreev levels, BCS-like theory for rough
interfaces between $d$-wave superconductors \cite{Grajcar02} predicts
\begin{equation}
(I_1R_N)_{\rm BCS}=(\pi/4)(\Delta/e)\cos 2\theta,
\label{eq:Walker} 
\end{equation}
where $\Delta$ is the maximal superconducting gap.  The measured
Josephson product of cuprate grain boundary junctions
\cite{Hilgenkamp02} can be well described by
$I_1R_N=\alpha^2(I_1R_N)_{\rm BCS}$ with a $\theta$-independent
renormalization factor $\alpha^2\sim 10^{-1}$.  In addition to the
$\theta>\theta_0$ data of \cite{Hilgenkamp02}, this functional form
can be tested also for $\theta=0$ (which case can be realized in break
junctions) with the result that average Josephson products of such
junctions \cite{Miyakawa99} are fully consistent with the grain
boundary data.  Moreover, in \cite{Miyakawa99} it has been shown that
$\Delta$ is not depressed in the junction region, thus explicitly
demonstrating the breakdown of the BCS prediction for $I_1R_N$ in the
cuprates.

There exists no generally accepted explanation of the small
renormalization factor $\alpha^2$. One of the reasons is that the
microstructure of Josephson junctions is typically quite complicated.
In fact, it is well known that small angle grain boundaries can be
modelled by a sequence of edge dislocations, while at larger
misorientation angles the dislocation cores start to overlap and no
universal picture applies to the structure of the grain boundary.  For
large-angle grain boundaries, Halbritter has proposed
\cite{Halbritter92} that the junction can be thought of as a nearly
impenetrable barrier with randomly placed highly conductive channels
across it. If due to strong Coulomb repulsion only the normal current
(and no supercurrent) is supported by these channels, the small value
of $I_1R_N$ follows quite naturally.

In this paper we shall argue that the smallness of the Josephson
product does not follow from the particular properties of the barrier,
but is rather an intrinsic property of the cuprates.  Such a point of
view has been first advocated in \cite{Deutscher99}.  However, that
paper did not consider alternative more conventional explanations.  In
order to support our point of view, let us begin by discussing the
Josephson product for intrinsic Josephson junctions in the $c$-axis
direction.  Such junctions can be viewed as an analogue of $ab$-plane
break junctions (since the misorientation angle vanishes for both),
but are preferable because of simpler geometry of the
interface. Moreover, zero energy surface bound states which may
develop at $ab$-plane surfaces because of the $d$-wave symmetry of the
pairing state \cite{Hu94} do not form in the $c$-axis direction,
simplifying the analysis of intrinsic Josephson junctions.

Standard BCS theory applied to the case of tunneling between
two-dimensional superconductors \cite{Bulaevskii73} predicts (for
coherent $c$-axis tunneling) that the $c$-axis critical Josephson
current density is $j_1=(2e/\hbar) N(0) \langle z_kt_k^2\rangle$,
where $\langle\ldots\rangle$ denotes an average along the (two
dimensional) Fermi line, $z_k$ is the wavefunction renormalization,
and $t_k$ is the matrix element for $c$-axis tunneling between
neighboring CuO$_2$ planes. $N(0)$ is the bare (unrenormalized)
density of states, $N(0)=\oint dk (4\pi^2\hbar v_k)^{-1}$, where the
integration runs along the two dimensional Fermi line and $v_k$ is the
bare Fermi velocity at the Fermi surface point $k$. Note that although
we have assumed that the self-energy is only frequency dependent, for
this geometry $z_k$ enters the expression for $j_1$ (and also the
normal-state resistance, see below).

Let us also note that the use of ordinary perturbation theory in
deriving an expression for $j_c$ has been criticized recently
\cite{Chakravarty98}. However, our formula yields the correct answer
for $t_k<\Delta$. This can be shown either by an explicit solution
of a $4\times 4$ Bogoliubov problem for two coupled planes with phase
differences 0 and $\pi$ between the planes, or by considering
solutions to the gap equation in an infinite layered system with a
finite $c$-axis total momentum of the Cooper pairs (for a similar
calculation, see e.g. \cite{Hlubina95}).

The conductance per square in the normal state is given by
$G_N=(2e^2/\hbar)N(0)\langle z_kt_k^2\Gamma_k^{-1}\rangle_N$, where
$\Gamma_k$ is the inverse lifetime of the quasiparticles and the index
$N$ in the Fermi line average means that the quantities are to be
evaluated in the normal state. Therefore standard theory predicts for
the Josephson product of intrinsic Josephson junctions
$I_1R_N=j_1G_N^{-1}\approx e^{-1}{\langle z_kt_k^2\rangle/ \langle
z_kt_k^2\Gamma_k^{-1}\rangle_N}$.  In conventional superconductors
$R_N$ can be measured at low temperatures in a sufficiently large
magnetic field. This is impossible for the cuprates and thus $R_N$ is
usually defined as the $c$-axis resistivity at $T_c$.

Unfortunately, due to the unknown temperature dependence of $z_k$, the
theoretical $I_1R_N$ can not be directly tested by experiment.  In
order to overcome this problem, in \cite{Latyshev99} instead of the
usual Josephson product a related characteristic of intrinsic
Josephson junctions has been studied, namely the product of the
critical current $I_1$ and of the resistivity $R_S$ in the resistive
mode of the junction (at low temperatures).  Since the conductance per
square in the resistive mode of the superconductor is
\cite{Latyshev99} $G_S\approx (8e^2/h) N(0) (z_kt_k^2)_{\rm
node}/\Delta$, standard BCS-like theory predicts
$I_1R_S=j_1G_S^{-1}\approx (\pi/2)(\Delta/e) {\langle z_k
t_k^2\rangle/(z_kt_k^2)_{\rm node}}$.  In \cite{Latyshev99},
$I_1R_S\sim \Delta/e$ has been found experimentally and good agreement
with theory has been claimed, since momentum-independent $t_k$ and
$z_k$ were assumed.  However, according to band structure calculations
\cite{Andersen94}, $t_k$ is strongly suppressed in the nodal
directions.  If this modulation of the tunnel matrix element $t_k$ is
taken into account and the presumably only moderate $k$-space
dependence of $z_k$ is neglected, the experimental $I_1R_S$ is seen to
be drastically reduced with respect to the theoretical predictions.

Thus we have shown that although the barriers in grain boundaries and
in intrinsic Josephson junctions are of very different nature, both
types of junctions exhibit a suppressed Josephson product. Therefore
we believe that this suppression is not due to specific barrier
properties as suggested in \cite{Halbritter92}, but rather due to some
intrinsic property of the high-$T_c$ superconductors.

\section{The second harmonic of the current-phase relation}

Since the second harmonic $I_2$ is not forced by symmetry to depend on
the angles $\theta_i$, its Josephson product can be estimated (for
temperatures larger than the energy of anomalous Andreev levels) using
the standard BCS theory as $I_2R_N\sim D\Delta/e$. Comparison with
Eq.~(\ref{eq:Walker}) implies that for junctions with $\theta\approx
45^\circ$ the $d$-wave symmetry of pairing leads to a suppression of
$I_1$, and $I_2$ may become comparable to $I_1$. This has in fact been
observed in two different types of 45$^\circ$ grain boundary Josephson
junctions \cite{Ilichev99,Ilichev01}.

However, the results of \cite{Ilichev99,Ilichev01} are quite
mysterious, if we take into account the actual experimental setup.  In
fact, standard BCS theory with ideal featureless barriers implies that
in order that $|I_2|\sim |I_1|$, the average misorientation angle
$\theta$ would have to be given with a precision $\sim D$, where
$D\sim 10^{-3}$.  This is not realistic and therefore two alternative
explanations have been proposed, in both of which the origin of the
anomalously large $|I_2/I_1|$ ratio has been sought in the barrier
properties.

{\it (i) Faceted scenario} has been considered as an alternative
explanation for symmetric $45^\circ$ junctions (i.e. junctions with
nominal geometry $\theta_1=0^\circ$ and $\theta_2=45^\circ$), in which
$|I_2|>|I_1|$ has been found \cite{Ilichev99}. It takes into account
the faceting of the grain boundary and also the twinned nature of the
(orthorhombic) YBCO thin films. Due to both of these features, the
junction can be viewed as a parallel set of 0 and $\pi$ junctions
\cite{Mannhart96}.  It has been shown \cite{Millis94} that in such a
case spontaneous currents are generated along the interface, the
ground state energy of the junction is minimized at a macroscopic
phase difference $\pm\pi/2$, and consequently the current-phase
relation is dominated by the second harmonic $I_2$.

In what follows we analyze quantitatively whether the faceted scenario
can apply to the results of \cite{Ilichev99}.  Let us denote the
current densities corresponding to the harmonics $I_i$ (with $i=1,2$)
as $j_i$ and introduce the Josephson penetration depth of the
junction, $\Lambda_J=(\Phi_0/4\pi\lambda\mu_0j_2)^{1/2}$.  Moreover,
let the local critical current density in the 0 and $\pi$ junctions be
$\pm j_0$, their typical length $a$, and the bulk penetration depth be
$\lambda$.  Then, since $a\approx 0.01-0.1 \mu$m, $\lambda\approx 0.15
\mu$m, and $\Lambda_J\sim 3\mu$m (estimated making use of
\cite{Hilgenkamp02} $j_2\sim 10^4$~A/cm$^2$), the inequalities
$\pi\lambda\gg a$ and $\Lambda_J^2\gg a\lambda$ are well satisfied.
Following \cite{Millis94} it is easy to show that these inequalities
guarantee that the spontaneously generated currents along the
Josephson junction can be calculated within perturbation theory.  If
we denote the total junction length by $L$, then a straightforward
calculation yields ${j_2/j_0}\approx \mu_0j_0a\lambda^2/\Phi_0$ and
${j_1/j_0}\approx (a/L)^{1/2}$.  The equation for $j_1$ is a random
walk-type formula, indicating that $j_1$ averages to zero in a
sufficiently long junction. After some algebra the above equations are
seen to imply ${j_2/j_1}\approx\sqrt{L\lambda/(4\pi\Lambda_J^2)}$.
Therefore, standard theory predicts that $I_2>I_1$ can be realized
only in sufficiently long junctions with $L>4\pi\Lambda_J^2/\lambda$.
This requires $L>500 \mu$m, whereas in \cite{Ilichev99} much shorter
junctions with $L\sim 1\mu$m were studied.

{\it (ii) Pinhole scenario} has been proposed in \cite{Ilichev01} as
an explanation for symmetric $45^\circ$ junctions (i.e.  junctions
nominally characterized by $\theta_2=-\theta_1=22.5^\circ$). It views
the barrier as basically impenetrable, the conduction being due to
randomly placed highly conductive pinholes. This explains quite
naturally the small value of the effective barrier transmission and,
at the same time, the large value of $|I_2/I_1|$.  Note that in order
to explain the small value of the Josephson product, in addition to
pinholes also Halbritter's conductive channels \cite{Halbritter92}
have to be postulated, which are assumed to be highly conductive only
in the normal and not in the superconducting channel.  Because of this
ad hoc nature of the pinhole scenario, and mainly because of the
absence of higher harmonics in $I(\phi)$ at 4~K \cite{Ilichev01} whose
presence it predicts, we believe that the pinhole picture should be
discarded.

Thus we conclude that the large value of the second harmonic $I_2$
(compared with predictions of the standard BCS theory) is most
probably not an extrinsic (barrier-related) effect, but rather an
intrinsic property of the cuprates.

\section{Microscopic implications}

The two apparently unrelated experimental facts, namely the suppressed
Josephson product $I_1R_N$ and the enhanced ratio $|I_2/I_1|$, can be
explained by a single assumption that in the cuprates some mechanism
is operative which leads to a suppression of $I_1$, while leaving
$R_N$ and $I_2$ intact. In what follows we describe one such mechanism
which we believe to be the most promising one.  Namely, we suggest
that at low temperatures the superconducting state of the cuprates
supports fluctuations of pairing symmetry towards $s$-wave pairing
(which pairing is expected to be locally stable within several
microscopic models of the cuprates).  Such fluctuations presumably do
not affect $R_N$, while they do influence the Josephson current. In
simplest terms, if we denote the phases of the superconducting grains
forming the junction as $\phi_i$, then the fluctuations renormalize
the first and second harmonics by the factors $\langle
e^{i\phi_1}\rangle\langle e^{i\phi_2}\rangle$ and $\langle
e^{2i\phi_1}\rangle\langle e^{2i\phi_2}\rangle$, respectively, where
$\langle\ldots\rangle$ denotes a ground-state expectation value.  Thus
experiment requires that the fluctuations have to be of such type that
$|\langle e^{i\phi}\rangle|=\alpha\approx 0.3$ and $|\langle
e^{2i\phi}\rangle| \approx 1$.  Precisely this behavior is expected if
the $d$-wave order parameter fluctuates towards $s$-wave pairing.

Now we proceed by introducing a minimal model of such
fluctuations. Unlike in standard literature on this subject (see,
e.g., \cite{Paramekanti00} and references therein), we assign a phase
field $\varphi_i$ to each bond $i$ of the square Cu lattice. In other
words, $\varphi_i$ lives on the sites of the dual lattice.  Since
large phase fluctuations are expected, the compactness of phase
fluctuations is explicitly taken into account and the model reads
\begin{eqnarray}
H=H_{\rm fluct}+
\sum_{\langle i,j\rangle}
\left[-V\cos(2\varphi_i-2\varphi_j)+W\cos(\varphi_i-\varphi_j)\right],
\label{eq:model}
\end{eqnarray}
where $H_{\rm fluct}=-m^{-1}\sum_i{\partial^2/\partial\varphi_i^2}$.
The second term in Eq.~(\ref{eq:model}) (with $\langle i,j\rangle$
denoting a pair of nearest neighbor sites) describes for $0<W<V$ a
superconductor with a dominant $d$-wave pairing $V+W$ and subdominant
$s$-wave pairing $V-W$. Since we concentrate on the ${\bf q}\approx
(\pi,\pi)$ fluctuations from $d$ to $s$-wave pairing, as a first
approximation we do not take into account the coupling of phase
fluctuations to electromagnetism.

In order to gain insight into the microscopics of $H_{\rm fluct}$, let
us recall that the RVB fluctuations in an elementary plaquette (in a
hole-free region) favor a minus sign between the two different valence
bond configurations (see Fig.~1). If these configurations are thought
of in terms of Bose condensates of valence bonds, this means that
energy is gained for a relative phase of the $x$ and $y$ condensates
$\arg(\sqrt{-1})=\pm \pi/2$. RVB processes are thus seen to frustrate
the phase ordering dictated by the second term in
Eq.~(\ref{eq:model}).

Since the other singlet state of the elementary plaquette (with a plus
sign between the valence bond configurations) lies higher in energy,
the effective Hamiltonian in the singlet sector is $H_{\rm
RVB}\propto\Delta^\dagger_i\Delta^\dagger_{i+x+y}\Delta_{i+x}\Delta_{i+y}+{\rm
H.C.}$, where $\Delta^\dagger_i$ creates a singlet on bond $i$.
Neglecting the amplitude fluctuations of $\Delta_i$, we can write
$H_{\rm
RVB}=J\sum_i\cos(\varphi_i-\varphi_{i+x}+\varphi_{i+x+y}-\varphi_{i+y})$.
In what follows we replace $H_{\rm fluct}$ by $H_{\rm RVB}$ in the
quantum model Eq.~(\ref{eq:model}), thereby obtaining a completely
classical toy model.

\begin{figure}[t]
\centerline{\includegraphics[width=4cm]{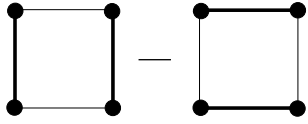}}
\caption{Ground state of the Heisenberg model on an elementary
plaquette.}
\end{figure}

\begin{figure}[t]
\centerline{\includegraphics[width=5cm]{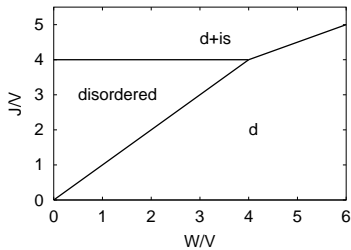}}
\caption{Mean field phase diagram of the toy model.}
\end{figure}

The toy model permits a simple mean field analysis: We consider two
types of solution, both of which live on two sublattices A and B, and
we explicitly disregard four-sublattice solutions, since they
correspond to translation-symmetry breaking states on the original Cu
lattice.  In the first type of solution $\varphi_A=\varphi$ and
$\varphi_B=0$ and there are two possibilities: either $\varphi=\pi$
($d$-wave), or $\varphi<\pi$ (a complex mixture of $d$ and $s$, to be
called $d+is$).  In the second type of solution we assume a disordered
state of such type that on sublattice $A$, the phase equals $0$ and
$\pi$ with probabilities $(1+\alpha)/2$ and $(1-\alpha)/2$,
respectively, and on sublattice $B$ the values $0$ and $\pi$ are
interchanged. In this case the macroscopic symmetry is of the $d$-wave
type, with renormalized averages $\langle
e^{i\varphi_A}\rangle=-\langle e^{i\varphi_B}\rangle=\alpha$ and
$\langle e^{2i\varphi_A}\rangle=\langle e^{2i\varphi_B}\rangle= 1$.
Minimization of energy (with respect to $\varphi$ or $\alpha$) leads
to the phase diagram shown in Fig.~2.  The renormalization factor in
the disordered $d$-wave state (which presumably corresponds to a
homogeneous but strongly fluctuating $d$-wave phase in the model
Eq.~(\ref{eq:model})) is $\alpha=(W/J)^{1/2}$. Thus the cuprates
correspond to the region $W\approx J/10$ and $J<4V$ in Fig.~2.

\section{Conclusions}
The anomalous effects observed in cuprate grain boundary and intrinsic
Josephson junctions can be explained by a single assumption of strong
quantum phase fluctuations at ${\bf q}=(\pi,\pi)$, which leads to
$|\langle e^{i\phi}\rangle|=\alpha\approx 0.3$ and $|\langle
e^{2i\phi}\rangle| \approx 1$.  In addition, our picture implies that
the Josephson product for junctions between the cuprates and low-$T_c$
superconductors is renormalized by the factor $\alpha$, in
semiquantitative agreement with experiment \cite{Sun96}. It also may
be relevant for the experiment \cite{Komissinski02}, where a large
second harmonic has been found in a $c$-axis Josephson junction
between YBCO and Nb.

\section{Acknowledgements}
I thank M. Grajcar for numerous stimulating discussions about the
Josephson effect.  I also thank T.~V.~Ramakrishnan for insightful
remarks on the $d$-$s$ phase fluctuations.  This work was supported by
the Slovak Scientific Grant Agency under Grant No.~VEGA-1/9177/02 and
by the Slovak Science and Technology Assistance Agency under Grant
No.~APVT-51-021602.

\end{document}